\documentclass[12pt]{iopart}

\begin{document}               

\def\be{\begin{equation}}
\def\ee{\end{equation}}
\def\ba{\begin{eqnarray}}
\def\ea{\end{eqnarray}}
\def\bas{\begin{eqnarray*}}
\def\eas{\end{eqnarray*}}
\newcommand{\ve}{\varepsilon}


\title{Level Repulsion in Constrained Gaussian Random--Matrix Ensembles}
\author{T.~Papenbrock$^{1,2}$, Z.~Pluha\v r$^3$, and H.~A.~Weidenm\"uller$^4$}
\address{$^1$ Department of Physics and Astronomy, University of Tennessee,
Knoxville, TN~37996, USA}
\address{$^2$ Physics Division,
Oak Ridge National Laboratory, Oak Ridge, TN 37831, USA}
\address{$^3$ Institute of Particle and Nuclear Physics, Faculty of
Mathematics and Physics, Charles University, 18000 Praha 8, Czech Republic}
\address{$^4$ Max-Planck Institut f\"ur Kernphysik,
D-69029 Heidelberg, Germany}

\begin{abstract}
Introducing sets of constraints, we define new classes of
random--matrix ensembles, the constrained Gaussian unitary (CGUE) and
the deformed Gaussian unitary (DGUE) ensembles. The latter interpolate
between the GUE and the CGUE. We derive a sufficient condition for
GUE--type level repulsion to persist in the presence of constraints.
For special classes of constraints, we extend this approach to the
orthogonal and to the symplectic ensembles. A generalized Fourier theorem
relates the spectral properties of the constraining ensembles with those
of the constrained ones. We find that in the DGUEs, level repulsion always
prevails at a sufficiently short distance and may be lifted only in the
limit of strictly enforced constraints.
\end{abstract}
\maketitle

\section{Introduction}

Ever since Wigner introduced random--matrix theory (RMT) in nuclear
physics in the 1950s~\cite{Wig56}, that theory has found wide
applications in modeling the fluctuation properties of spectra and wave
functions of complex systems ranging from vibrating crystals, to
microwave resonators, to quantum dots, to atoms, and to the Dirac
operator in lattice QCD~\cite{Bro81,GMW98}. Depending on the symmetry of
the Hamiltonian, there are three ``classical'' random--matrix ensembles:
the Gaussian orthogonal ensemble (GOE), Gaussian unitary ensemble (GUE),
and Gaussian symplectic ensemble (GSE) which consist of real symmetric,
complex Hermitean, and complex quaternion matrices, respectively. The
corresponding matrix elements are independent, Gaussian-distributed
random variables with $\beta = 1,2, \ \mbox{and} \ 4$ degrees of
freedom, respectively. One of the hallmarks of RMT is level repulsion,
i.e., the probability of finding a spacing $s$ between two closely
spaced neighboring levels is proportional to $s^\beta$.

Many complex physical systems exhibit spectral fluctuation properties
that are in agreement with those of the Gaussian random-matrix ensembles,
although the corresponding Hamiltonian matrices have structures that
differ considerably from a Gaussian random matrix. Many--body systems
with $k$--body interactions, for instance, have sparse Hamiltonian
matrices, since many--body states that differ in the occupation of
more than $k$ single--particle orbitals cannot be connected by the
interaction. Only a few analytical results are known for these
$k$--body embedded Gaussian ensembles \cite{Mon75}, and we refer the
reader to the recent reviews~\cite{Kota01,Ben03}. Realistic
random--matrix models for nuclei and atoms also have to include spin
and isospin symmetry. The resulting two--body random ensemble (TBRE)
\cite{FreWon,BohFlo} is mathematically very complicated, and virtually
no analytical results are known regarding the fluctuation properties
of this important random--matrix model~\cite{Pap06}. Another example
is given by quasi one--dimensional disordered electronic systems. The
corresponding Hamiltonians are band matrices with zero elements
outside a small band around the diagonal. The theoretical description
of random band matrices is possible due to the simple structure of the
Hamiltonian~\cite{Fyo91}.

The embedded Gaussian $k$--body ensembles, the TBRE, and the ensembles
of random band matrices have one property in common. They can be
viewed as {\it constrained} random-matrix ensembles. Because of the wide
occurrence of such matrix ensembles in various branches of theoretical
physics, the understanding of their spectral fluctuation properties
offers a considerable challenge to theory. So far, the available
evidence (mostly based on numerical simulations for matrices of rather
small dimensions) points towards fluctuation properties of standard RMT
type. The present paper is a first step toward a common theoretical
treatment of constrained ensembles. These ensembles are introduced in
Section~\ref{CGUE}, and are defined as Gaussian ensembles of matrices
where certain linear combinations of matrix elements vanish. We note
that the GOE can ultimately also be viewed as a constrained ensemble;
it is obtained from the GUE through the constraint of vanishing
imaginary parts of all off--diagonal matrix elements. Remarkably, some
properties of constrained Gaussian random-matrix ensembles do not depend
on the details of the constraints but only on their number and symmetry
properties (as encoded in systematic degeneracies). In this work, we
focus mainly on these rather general properties.

This article is organized as follows. Section~\ref{CGUE} introduces
constrained Gaussian random matrix ensembles. In Section~\ref{S_HCIZ},
we employ the Harish--Chandra Itzykson Zuber (HCIZ) integral formula
to express the joint probability function for the eigenvalues of the
constrained ensemble through an integral over the constraints. In
Section~\ref{suff}, we derive a sufficient condition for the existence
of quadratic, GUE--like level repulsion. The condition is formulated
as a simple inequality that relates the number of constraints and the
number of degeneracies to the dimension of Hilbert space. We
illustrate this result with a number of examples. In
Section~\ref{newsec} we analyse the spectal width and the distribution
of matrix element of the CGUE. In Section~\ref{FT}, we show that the
joint probability function of the constrained ensemble is related to
that of the constraining ensemble by a generalized Fourier transform.
In Section~\ref{deform}, we consider {\it deformed} matrix
ensembles. These ensembles interpolate between the GUE and the
constrained ensembles and allow us to study situations where the
constraints are slowly switched on. We conclude with a Summary.

\section{Constrained Gaussian unitary random--matrix ensembles}
\label{CGUE}

We consider Hermitean matrices acting on a Hilbert space ${\cal H}$ of
dimension $N$. Together with ${\cal H}$, we also consider the linear
space ${\cal V}$ spanned by the Hermitean matrices acting on ${\cal
H}$: Every linear combination $a A + b B$ of two such matrices $A, B$
with real coefficients $a, b$ is also a Hermitean matrix acting on
${\cal H}$. In the linear space ${\cal V}$, we introduce the canonical
scalar product in terms of the trace
\be
\langle A |B \rangle \equiv {\rm Tr} (A B)
\label{scalar}
\ee
for every pair $A, B$ of matrices. This allows us to define an
orthonormal basis of $N^2$ Hermitean basis matrices $B_\alpha =
B^{\dagger}_\alpha$ in ${\cal V}$ which obey
\be
\label{basis}
\langle B_\alpha|B_\beta\rangle \equiv {\rm Tr} (B_\alpha B_\beta) 
= \delta_{\alpha \beta}
\ee
and
\be
\sum_{\alpha = 1}^{N^2} | B_\alpha \rangle \langle B_\alpha | =
{\bf 1} \ ,
\label{compl}
\ee
where ${\bf 1}$ is the unit operator in ${\cal V}$. Such a set of
basis matrices is, for instance, given by matrices that have a unit
matrix element somewhere in the main diagonal and zeros everywhere
else, or an element $(1/\sqrt{2})$ somewhere above the main diagonal
and its mirror image below the main diagonal and zeros everywhere else,
or an element $(i/\sqrt{2})$ somewhere above the main diagonal and its
complex conjugate in the mirrored position below the main diagonal and
zeros everywhere else. This choice of a basis is but one example. Any
other basis obtained by orthogonal transformations in ${\cal V}$ (with
an orthogonal matrix of dimension $N^2$) from the one just described is
equally admissible. In general we have no preference in this respect.

Any Hermitean matrix $H$ acting on ${\cal H}$ can be expanded in terms of
the $N^2$ Hermitean basis matrices $B_\alpha$ as 
\be
\label{expan}
H=\sum_{\alpha=1}^{N^2} h_\alpha B_\alpha.
\ee
Thus, $H$ can also be viewed as a vector in ${\cal V}$.

We consider the decomposition of ${\cal V}$ into two orthogonal
subspaces labeled ${\cal P}$ and ${\cal Q}$. The decomposition is
defined in terms of orthogonal projection operators
\ba
{\cal P} &=& \sum_{p =1}^{N_P} | B_p \rangle \langle B_p
| \ , \nonumber \\
{\cal Q} &=& \sum_{q = N_P + 1}^{N^2} | B_q \rangle \langle B_q
| \ .
\label{A3}
\ea
We have
\be {\cal P}^{\dagger} = {\cal P} \ , \ {\cal Q}^{\dagger} =
{\cal Q} \ , \ {\cal P}^2 = {\cal P} \ , \ {\cal Q}^2 = {\cal Q} \ , \
{\cal P} {\cal Q} = 0 \ , \ {\cal P} + {\cal Q} = {\bf 1} \ .
\label{A4}
\ee
When we sum over matrices in ${\cal P}$--space or in ${\cal Q}$--space,
we will sometimes use the summation indices $p$ and $q$, respectively,
without referring to the partition explicitly while a summation over
all $N^2$ basis matrices is indicated by Greek summation indices. The
operators ${\cal P}$ and ${\cal Q}$ are defined in ${\cal V}$ and have
dimension $N_P$ and $N_Q = N^2 - N_P$, respectively. We have been
rather explicit in the construction of ${\cal P}$ and ${\cal Q}$
because these operators differ from the projection operators often used
in Hilbert space: The latter are defined in terms of the basis vectors
spanning Hilbert space, while our projection operators are defined in
terms of the basis matrices $B_\alpha$ and operate in ${\cal V}$. 

In this article, we study constrained ensembles of Gaussian random
matrices. Every member $H$ of the ensemble is constrained to have
zero projection onto ${\cal Q}$--space 
\be 
{\cal Q} H = 0 \ {\rm or} \ {\cal P} H = H \ . 
\label{A5} 
\ee 
This condition can also be expressed as 
\be 
\langle B_q | H \rangle = 0 \ {\rm for \ all} \ q \ {\rm or} \ H =
\sum_p B_p \langle B_p | H \rangle \ . \label{A6}
\ee 
We combine condition~(\ref{A6}) with the assumption that the ensemble
has a Gaussian distribution. Using the constraint in the form of the
first of Eqs.~(\ref{A6}), we write the probability density $\tilde{W}
_{\cal P}(H)$ as
\be
\tilde{W}_{\cal P}(H) {\rm d} [ H ] = (2\pi)^{-N_P/2} 
\exp{\left( -{1\over 2} \langle H | H\rangle  \right)} 
\prod_q \delta ( \langle B_q | H \rangle ) \ {\rm d} [ H ] \ .
\label{A7a}
\ee
Here ${\rm d} [ H ]$ stands for the product of the differentials
of the $N^2$ independent matrix elements of $H$. Eq.~(\ref{A7a})
defines the constrained Gaussian ensemble of random matrices for the
set $\{B_q\}$ of constraining matrices.

Every unitary transformation $U$ in Hilbert space induces a
transformation $B_q \to U B_q U^{\dagger}$ of the set of matrices
defining ${\cal Q}$--space. We say that the set $\{ B_q \}$ is
invariant under unitary transformations if the vector spaces spanned
by the sets $\{ B_q \}$ and $ \{ U B_q U^{\dagger} \} $ are
identical. (This is not the case in general). Let us consider an
arbitrary unitary transformation $H \to U H U^{\dagger}$ of the
matrices in the ensemble~(\ref{A7a}). All terms on the right--hand
side of Eq.~(\ref{A7a}) remain unchanged except for the arguments of
the delta functions. Here, the unitary transformation is tantamount
to replacing $B_q \to U B_q U^{\dagger}$. Therefore, $\tilde{W}_{\cal
P}$ is unitarily invariant if and only if the set $\{ B_q \}$ has
this invariance property. Lack of unitary invariance implies, in
general, that the eigenvalues and eigenvectors are correlated random
variables.

This is problematic only at first sight. Indeed, the transformation
$\{B_q\} \to \{ U B_q U^{\dagger}\}$ is exactly compensated by the
transformation $\{H = \sum_p h_p B_p\} \to \{U H U^{\dagger} = \sum_p
h_p U B_p U^{\dagger}\}$. Instead of considering the sets of
constraints $\{B_q\}$ and $\{ U B_q U^{\dagger}\}$, we may consider the
given set $\{B_q\}$ and two constrained ensembles, one comprising the
matrices $\{H\}$ as defined by Eq.~(\ref{A7a}) and the other, all
matrices obtained from that set by the operation $H \to U H U^{\dagger}$,
with $U$ fixed. The matrices $H$ and $U H U^{\dagger}$ possess identical
spectra, and in each ensemble the corresponding sets of eigenvalues are
encountered with equal probability. Therefore, the two ensembles can
differ only in the ordering of the eigenvalues, i.e., by permutations.
Such a difference can arise only if the eigenvalue distribution of the
original ensemble $\tilde{W}_{\cal P}$ is not symmetric.

We group the sets $\{B_q\}$ of constraining matrices into equivalence
classes. Together with a given set $\{ B_q \}$, each class comprises all
sets $ \{ U B_q U^{\dagger} \}$ obtained from $\{ B_q \}$ by arbitrary
unitary transformations $U$ in $N$ dimensions.  The ``superensemble''
$W$ is the union of all ensembles $\tilde{W}$ with constraints in the
same equivalence class. By construction, $W$ is unitarily invariant. We
refer to $W$ as to the constrained Gaussian unitary random--matrix
ensemble (CGUE). The CGUE is ergodic if the members of the equivalence
class possess symmetric eigenvalue distributions (i.e., distributions
that are invariant under all permutations of the eigenvalues). In this
case, any result derived for the superensemble also holds for each
of its member ensembles. An example is given by the GOE as obtained by
constraining the GUE. The GOE is invariant under orthogonal
transformations, and the eigenvalue distribution must, therefore, be
totally symmetric. Therefore, the superensemble is ergodic. A
counterexample is given by the constraint that $H$ have zero
non--diagonal matrix elements in the first row and column. The eigenvalue
distribution factorizes and is not symmetric. The constraint is unitarily
equivalent, among many others, to the constraint that $H$ have zero
non--diagonal matrix elements in the $k$th row and column, with $k = 2,
\ldots, N$. Thus, results derived for the spectral fluctuation properties
of the associated CGUE apply jointly to all these constrained ensembles
but not to the individual members. We see that our focus on the CGUE
prohibits the study of unsymmetrical eigenvalue distribution functions
which might be useful, e.g., in studies of symmetry-breaking.
It may happen that the vector space spanned by the set $\{B_q\}$ is
invariant under a subgroup of the unitary group. In that case, the
equivalence class is generated by the coset space of the unitary group
with respect to that subgroup.

For a set of constraints $\{B_q\}$, the CGUE is defined by
\ba
W_{\cal P}(H) {\rm d} [ H ] = \nonumber\\
(2\pi)^{-N_P/2} \exp{\left( -{1\over 2}
\langle H | H\rangle  \right)} {\rm d} [ H ] \int {\rm d} [U] \bigg(
\prod_q \ \delta ( \langle U B_q U^{\dagger} | H \rangle ) \bigg) \ .
\label{A7b}
\ea
The integral ${\rm d}[U]$ extends over the unitary group in $N$ dimensions.
The ensemble is obviously invariant under unitary transformations of $H$.
The Haar measure of the unitary group is normalized to one, i.e.
\be
\int {\rm d} [U] = 1 \ .
\ee
Proceeding as usual, we diagonalize the matrix $H$ with the help of a
unitary matrix $V$,
\be
H = V x V^{\dagger} \ ,
\label{A8}
\ee
where $x={\rm diag}(x_1,\ldots,x_N)$ is the diagonal matrix of the
eigenvalues. Here, and in what follows, we denote diagonal matrices by
small letters. The integration measure becomes
\be
{\rm d} [H] = {(2\pi)^{N(N-1)/2}\over \prod_{k = 1}^N k!}
\Delta^2(x) {\rm d} [x] {\rm d} [V] \ ,
\label{dh}
\ee
where $\Delta(x)$ denotes the Vandermonde determinant
\be
\Delta(x) = \prod_{1\le j<k\le N} (x_k-x_j) \ .
\ee
Eq.~(\ref{dh}) shows that eigenvalues and eigenvectors are uncorrelated
random variables. The distribution of the eigenvectors is defined by the
Haar measure. Therefore, the joint probability distribution $P_{\cal
P}(x)$ of the eigenvalues is the object of central interest in this
paper. It is given by
\be 
P_{\cal P}(x) = {(2\pi)^{(N_Q - N)/2}\over \prod_{k = 1}^N k!}
\exp{\left( - {1\over 2}\langle x | x \rangle \right)} \Delta^2(x)
F_{\cal P}(x) \ ,
\label{A8a} 
\ee
where
\be
F_{\cal P}(x) \equiv \int {\rm d} [ U ] \bigg( \prod_q 
\delta ( \langle B_q | U x U^{\dagger} \rangle ) \bigg) \ .
\label{A9a}
\ee

This construction of the CGUE may seem unnecessarily involved. Why
not start from the constrained ensemble defined in Eq.~(\ref{A7a}),
use the transformation of variables and measures as given by
Eq.~(\ref{dh}), and integrate $\tilde{W}$ over the unitary group to
obtain the distribution function for the eigenvalues? Following this
path, we actually arrive at Eqs.~(\ref{A8a}) and (\ref{A9a}) and, thus,
at the CGUE: In this approach, the eigenvalue distributions are always
symmetric under permutations of the eigenvalues, since the integration
over the unitary group sums over all $N!$ different unitary matrices
that diagonalize a given matrix $H$ with $N$ non-degenerate
eigenvalues. We believe that our line of reasoning shows more
clearly the conceptual framework we use.

It would be desirable to have a classification scheme for the CGUEs.
How many such ensembles are there for a given number of constraints?
Equivalently, how many different equivalence classes exist for a fixed
number of constraining matrices? We have not investigated these
questions yet. In the present paper, we focus attention on joint
properties of the CGUEs.

Because of the presence of the function $F_{\cal P}(x)$ on the
right--hand side of Eq.~(\ref{A8a}), the spectral statistics of the
CGUE differ from those of the GUE. To work out that difference, we
must study the function $F_{\cal P}(x)$.

\section{The HCIZ Integral}
\label{S_HCIZ}

To calculate $F_{\cal P}(x)$, we replace the delta functions by Fourier
integrals, introducing the set $\{ t_q \}$ of $N_Q$ integration
variables. Then,
\be
F_{\cal P}(x) = (2 \pi)^{-N_Q} \int \prod_q {\rm d} t_q
\int {\rm d} [U]  \exp{\left( i \sum_q t_q
\langle B_q | UxU^\dagger \rangle \right)} \ .
\label{A9b}
\ee
The integral over the unitary group is the
Harish--Chandra Itzykson Zuber (HCIZ) integral~\cite{HC57,IZ80,zinn}.
It can be expressed in terms of the eigenvalues $b_j(\{ t_q \})$, $j =
1, \ldots, N$ of the matrix
\be
B(\{t_q\}) = \sum_q t_q B_q \ ,
\label{A9d}
\ee
and is given by 
\be
\label{hciz}
\int d[U] \exp{\left(i\langle B|UxU^\dagger\rangle\right)} =
\left(\prod_{j=1}^{N-1}j!\right) i^{-N(N-1)/2} 
{\det [\exp{(i x_k b_l)}] \over \Delta(x)\Delta(b)}.
\ee
The determinant has $x_k$ in the $k$th row, and $b_l$ in the $l$th
column, with $k,l = 1, \ldots, N$. We thus obtain
\be
F_{\cal P}(x) = {\prod_{j = 1}^{N-1} j! \over (2\pi)^{N_Q} \
i^{N(N-1)/2}} \int \prod_q {\rm d} t_q \frac{\det [ \exp{( i x_k b_l )}]}
{\Delta(x) \Delta(b) }
\ .  
\label{A9c}
\ee
The function $F_{\cal P}(x)$ is obviously symmetric with respect to
all permutations of the eigenvalues $x_k$. The integrand 
of $F_{\cal P}(x)$ in
Eq.~(\ref{A9c}) has that same property with respect to the
eigenvalues $b_l$ of the matrix $B$. Both properties are a
consequence of the unitary invariance of the CGUE. Multiplying every
eigenvalue $x_k$ by a parameter $\tau$ so that $x_k \to \tau x_k$ for
all $k$, we get
\be
F_{\cal P}(\tau x) = \tau^{- N_Q} F_{\cal P}(x) \ .
\label{A11a}
\ee
This shows that $F_{\cal P}(x)$ is a homogeneous function of the
eigenvalues which has a singularity of order $N_Q$ when all
eigenvalues tend to zero simultaneously. However, $F_{\cal P}(x)$
does not diverge generically when any one of the eigenvalues vanishes
individually.

Using Eq.~(\ref{A9c}) in Eq.~(\ref{A8a}), we find for the joint
probability distribution of the eigenvalues the expression
\be 
\label{P_Pt}
P_{\cal P}(x) = 
\frac{\exp{\left( - {1\over 2}\langle x | x \rangle\right)}\Delta^2(x)}
{(2\pi)^{(N + N_Q)/2} \ i^{N(N-1)/2}N!}
\int \prod_q {\rm d} t_q 
\frac{\det [\exp{(i b_j x_k )}]}{\Delta(b)\Delta(x)} \ .
\label{A10} 
\ee

\section{Sufficient condition for level repulsion}
\label{suff}

\subsection{Unitary case}

We use the function $F_{\cal P}(x)$, as given in Eq.~(\ref{A9c}), to
investigate the spectral statistics and, in particular, level
repulsion at short distances for the constrained ensembles. It is
obvious that GUE--like level repulsion, as given by the factor
$\Delta^2(x)$ in Eq.~(\ref{A10}), will prevail unless the function
$F_{\cal P}(x)$ has singularities whenever two eigenvalues $x_j, x_k$
coincide. This prompts us to study especially the singularities of
$F_{\cal P}(x)$.

The integrand in Eq.~(\ref{A9c}) has no singularities: Any zeros of
the denominator are canceled by corresponding zeros of the numerator.
Indeed, zeros of the denominator arise when two or more eigenvalues
$b_j$ of the matrix $B$ coincide. We assume that $b_1 = b_2 = \ldots
= b_L$ are $L$ degenerate eigenvalues. Without loss of generality, we
may assume that they vanish. (Under translation $b_j \to b_j + \Delta
b$ with fixed $\Delta b$ for all $j$, the integrand is multiplied by
the non--singular function $\exp{( i \Delta b \sum x_k )}$). Then the
integrand becomes
\ba
\label{int_degen}
\fl
\qquad
\frac{\det[\exp{( i b_j x_k )}]} {\Delta(x) \Delta(b) }
\bigg|_{b_1=\ldots=b_L=0} 
= \\\nonumber
\left(\prod_{l=1}^{L-1}{i^l\over l!}\right)
\frac{\det \left[ \begin {array}{ccccccc} 
1 & x_1 &\ldots& x_1^{L-1}& e^{ix_1b_{L+1}} &\ldots & e^{ix_1 b_N}\\
1 & x_2 &\ldots& x_2^{L-1}& e^{ix_2b_{L+1}} &\ldots & e^{ix_2 b_N}\\
\vdots & \vdots & &\vdots &\vdots& &\vdots \\
1 & x_N &\ldots& x_N^{L-1}& e^{ix_Nb_{L+1}} &\ldots & e^{ix_N b_N}
\end{array}\right]} 
{\left(\prod_{n=L+1}^N b_n\right)^L \Delta(b_{L+1},\ldots,b_N)
\Delta(x)} \ ,
\ea
(where $\Delta(b_{L+1},\ldots,b_N)$ denotes the Vandermonde determinant
of the eigenvalues $b_{L+1}, \ldots, b_N$) and is obviously finite. The
same conclusion applies when two or more eigenvalues $x_k$ coincide. For
a discussion and alternative calculation of the HCIZ integral in the
presence of degeneracies, we refer the reader to Refs.~\cite{Sch03,Fyo03}.

We conclude that the $N_Q$--dimensional integral in Eq.~(\ref{A9c}) 
can diverge only because the domain of integration is not bounded. This
suggests that we introduce $N_Q$--dimensional spherical coordinates
$\prod_q {\rm d}t_q \equiv t^{N_Q - 1} {\rm d}t {\rm d}\Omega$ (with $t$
the radial variable and $\Omega$ the set of angular variables), and
focus attention on the radial integral
\be
\int\limits_0^\infty {\rm d}t \ t^{N_Q - 1} \frac{\det[\exp{( i b_j x_k
)}]}{\Delta(b) } \ .
\ee
From Eq.~(\ref{A9d}), it follows that the eigenvalues are linear
homogeneous functions of the variables $t_q$ so that $b_j(\{\lambda
t_q \}) = \lambda b_j(\{ t_q \})$. This implies $b_j(t,\Omega) = t
b_j(1,\Omega)$.

We distinguish two cases. (i) There are no {\it systematic} degeneracies
among the eigenvalues $b_j$, i.e., there exists no domain of integration
with measure $> 0$ in which eigenvalues are degenerate. (Accidental
degeneracies that occur on a set of measure zero of the integration
domain are irrelevant.) Then, the Vandermonde determinant in the
denominator yields a factor $t^{N(N-1)/2}$ and the radial integral
becomes
\be
\int\limits_0^\infty {\rm d}t \ t^{N_Q - 1 - N(N-1)/2} 
\det [\exp{( i t b_j(1,\Omega) x_k)}] \ .
\ee
The generalization of Eq.~(\ref{int_degen}) to $L = N$ degenerate
eigenvalues shows that the integrand is not singular at $t=0$. The
radial integral is sure to converge if $N_Q < N(N-1)/2$. (ii) There is a
set of $L$ systematically degenerate eigenvalues. The domain of
degeneracy extends out to $t = \infty$. In this case, we use the
integrand in the form~(\ref{int_degen}). For the radial integral,
this yields
\be
\fl
\quad
\int\limits_0^\infty {\rm d}t \ t^{N_Q-1+L(L-1)/2-N(N-1)/2 }
\det \left[ \begin {array}{ccccccc} 
1 & x_1 &\ldots& x_1^{L-1}& e^{ix_1b_{L+1}} &\ldots & e^{ix_1 b_N}\\
1 & x_2 &\ldots& x_2^{L-1}& e^{ix_2b_{L+1}} &\ldots & e^{ix_2 b_N}\\
\vdots & \vdots & &\vdots &\vdots& &\vdots \\
1 & x_N &\ldots& x_N^{L-1}& e^{ix_Nb_{L+1}} &\ldots & e^{ix_N b_N}
\end{array}\right] \ .
\ee
Again, the integrand is well behaved at $t = 0$, and the integral 
is guaranteed to converge for $N_Q + L(L-1)/2 < N(N-1)/2$. This result
can be generalized. Let there be $J$ sets of $L_j, j = 1, \ldots,
J$ degenerate eigenvalues with degeneracy domains that extend out
to $t = \infty$. Then, quadratic GUE--like level repulsion is
guaranteed to occur if the number of constraints $N_Q$, the numbers
of degenerate levels $L_j$, and the matrix dimension $N$ obey the
inequality
\be
\label{main}
N_Q + \sum_{j=1}^J L_j(L_j-1)/2 < N(N-1)/2.
\ee
The condition~(\ref{main}) is a sufficient (but not neccessary)
condition for GUE--type level repulsion to occur in the presence
of $N_Q$ constraints. (It may happen that the exponential functions
in the integrand provide convergence even if condition~(\ref{main})
is violated.) The inequality~(\ref{main}) makes no reference to the
specific structure of the constraints; the only input is the number
of constraints and the number of degeneracies. We recall that
degeneracies are related to symmetries. In this sense, the number and
symmetry properties of the constraints are at the root of the
inequality~(\ref{main}).

The possibility to derive the general result~(\ref{main}) without
being specific about the nature of the constraints is, of course, due
to the unitary symmetry of the CGUE. We have mentioned already that
the function $F_{\cal P}(x)$ will affect spectral fluctuation
properties not only on a very short scale (level repulsion) but also
on larger scales. It appears, however, that that influence depends on
specific properties of the constraints and requires explicit
calculation of $F_{\cal P}(x)$.

\subsection{Orthogonal and symplectic case}

Instead of considering the Hilbert space of Hermitean matrices, we
might have started out from the Hilbert spaces of real symmetric or
complex--quaternion matrices. In both cases, we follow the same
steps as for the CGUE and arrive at the analogue of Eq.~(\ref{A8a})
\be 
P_{\cal P}(x) \propto 
\exp{\left( - {1\over 2}\langle x | x \rangle \right)} 
|\Delta(x)|^\beta F_{\cal P}(x)\ .
\label{Pbeta} 
\ee 
The factor $|\Delta(x)|^\beta$ stems from the Jacobian of the
eigenvalue/eigenvector transformation for real symmetric ($\beta = 1$),
and complex-quaternion matrices ($\beta = 4$). We have omitted the
overall normalization constant. Likewise, we have not distinguished the
functions $P_{\cal P}$ and $F_{\cal P}$ from their unitary counterparts 
as there is no room for confusion. The function $F_{\cal P}$ is 
analogous to Eq.~(\ref{A9b}) and reads
\be
F_{\cal P}(x) \propto \int \prod_q {\rm d} t_q
\int {\rm d} [U_\beta]  \exp{\left( i \sum_q t_q
\langle B_q | U_\beta x U_\beta^\dagger \rangle \right)} \ .
\label{Fbeta}
\ee
We denote orthogonal and symplectic matrices as $U_1$ and $U_4$,
respectively. The integral ${\rm d}[U_\beta]$ is over the orthogonal
($\beta = 1$) or symplectic ($\beta = 4$) group in $N$ dimensions. The
constraining matrices $B_q$ are real and symmetric for $\beta = 1$ and
complex quaternion for $\beta = 4$. Eqs.~(\ref{Pbeta}) and (\ref{Fbeta})
define the spectral statistics for the constrained Gaussian orthogonal
ensemble (CGOE) and the constrained Gaussian symplectic ensemble (CGSE).

Progress is hampered by the fact that the analogue of the HCIZ
integral is not available in closed form for the orthogonal and the
symplectic groups. However, we might employ an asymptotic expansion of
the HCIZ integral that has recently been re--derived in Ref.~\cite{Guh02}.
We assume that neither the eigenvalues $x$ nor the eigenvalues of the
matrix $B$ possess any systematic degeneracies. Again, we introduce
spherical coordinates $\prod_q {\rm d}t_q \equiv t^{N_Q - 1} {\rm d}t
{\rm d}\Omega$. We are interested in the behavior of the integrand for
large values of the radial coordinate $t$. Since $b_i(t,\Omega)= tb_i(1,
\Omega)$, the differences of eigenvalues of $B$ grow linearly with $t$,
and we may confine ourselves to the leading term in the asymptotic
expansion
\be
\int {\rm d} [U_\beta]  \exp{\left( i \sum_q t_q
\langle B_q | U_\beta x U_\beta^\dagger \rangle \right)} \approx 
\frac{\det[\exp{( i b_j x_k )}]} {|\Delta(x) \Delta(b)|^{\beta/2} } \ . 
\ee
Correction terms are inversely proportional to powers of the differences
of the eigenvalues $(b_j - b_k)$. Following the same arguments as in the
unitary case, we find that the radial $t$--integration is guaranteed to
converge if 
\be
\label{mainbeta}
N_Q < \beta N(N-1)/4 \ . 
\ee
In the absence of any systematic degeneracies of the constraints, the
inequality (\ref{mainbeta}) is a sufficient condition for the existence
of eigenvalue repulsion of the canonical form $|x_i-x_j|^\beta$ in the
constrained Gaussian ensembles with Dyson index $\beta$. The more general
result~(\ref{main}) for $\beta = 2$ and the inequality~(\ref{mainbeta})
are central results of our work.

\subsection{Examples}

We present several examples for the CGUE, the CGSE, and the CGOE, which
illustrate the power and limitations of the inequalities derived above.

(i) We consider the GOE as obtained by constraining the GUE. The ${\cal
P}$--space consists of real symmetric matrices and the ${\cal Q}$--space
of purely imaginary antisymmetric matrices. We have $N_Q = N(N - 1)/2$
and no systematic degeneracies. The condition~(\ref{main}) is violated,
as must be the case, because the eigenvalue distribution function of the
GOE carries $|\Delta(x)|$ in the first power. This implies that
\be
F_{\cal P}(x) \propto \frac{1}{|\Delta(x)|} \ .
\label{iden1}
\ee
This relation can be proved by employing a relation between complementary
distribution functions that is derived in Section~\ref{FT}; see also the
Appendix.

(ii) We consider the case where ${\cal Q}$--space is spanned by the
set of diagonal matrices. The number of constraints is $N$ and the
condition~(\ref{main}) is fulfilled. Level repulsion prevails. For $N
\gg 1$, the corrections to the GUE average level density and the
two--point function can be worked out analytically. We use the
supersymmetry technique and the standard saddle--point approximation.
This approximation consists in neglecting terms which are small of order
$1/N$ and results in an asymptotic expansion in inverse powers of $N$.
The deviations from the GUE caused by the constraints are of order $1/N$
and, therefore, do not affect the saddle--point condition. Hence, we find
that the corrections to the GUE results vanish like inverse powers of $N$.

(iii) This case is converse to the previous one: ${\cal P}$--space is
spanned by the set of diagonal matrices. The number of constraints is
$N_Q = N^2 - N$, and condition~(\ref{main}) is violated. The levels in
${\cal P}$--space obey Poisson statistics. Therefore, the function
$F_{\cal P}$ must compensate the factor $\Delta^2(x)$ in Eq.~(\ref{A10}).
Hence we must have
\be
F_{\cal P}(x) \propto \frac{1}{\Delta^2(x)} \ .
\label{iden2}
\ee
It seems difficult to prove the relation~(\ref{iden2}) by direct
calculation. 

(iv) Block-diagonal Hermitean matrices: We require that two rectangular
blocks in $H$ carry zero matrix elements. If the two blocks consist of all
non--diagonal matrix elements in row $k$ and in column $k$, the number of
constraints is $2 (N - 1)$. In this case, $N - 2$ of the eigenvalues of
$B$ are degenerate (i.e., vanish), and the remaining two eigenvalues
differ only in sign. The condition~(\ref{main}) is violated (albeit weakly).
These considerations are easily extended to bigger blocks. 

(v) Another example is furnished by the EGUE, the embedded Gaussian
unitary ensemble with $k$--body interactions~\cite{Mon75,Kota01,Ben03}.
This ensemble differs from a CGUE in that its matrices $B_p$ do not obey
the normalization~(\ref{basis}). However, the corresponding CGUE is dense
in the EGUE. With $l$ the number of degenerate single--particle states, and
$m$ the number of identical Fermions, we have $N = { l \choose m }$. The
number of independent $k$--body matrices $B_p$ is $N_P = 2({ l \choose k})^2
- {l\choose k}$. According to condition~(\ref{main}), level repulsion is
guaranteed if there are no degeneracies and if $N(N + 1)/2 < N_P$, or if
\be
{1\over 2}{ l \choose m }\left({ l \choose m } + 1\right)
< { l \choose k}\left(2{l\choose k} -1\right) \ .
\ee
For arbitrary $l$, this condition is met only for $k = m$, and there it is 
expected. Unfortunately, we cannot draw any non--trivial conclusions for
the EGUE.

(vi) ${\cal P}$-space consists of all traceless Hermitean matrices. The
single constraint ($N_Q = 1$) is proportional to the unit matrix, and all
$N$ eigenvalues of the constraint are degenerate. The probability
distribution~(\ref{P_Pt}) can be worked out by means of
Eq.~(\ref{int_degen}), and the result differs from the GUE eigenvalue
distribution only by an overall factor $\delta(x_1+x_1+\cdots +x_N)$,
which yields a GUE spectrum with vanishing centroid. The spectral
fluctuations are identical to the GUE though inequality~(\ref{main}) is
not fulfilled. This demonstrates that the conditions~(\ref{main}) and,
by the same token, (\ref{mainbeta}) are only sufficient: Their fulfillment
guarantees level repulsion with power $\beta$, and they cannot be
fulfilled if there is no level repulsion with power $\beta$. However, no
conclusion can be drawn about level repulsion if these inequalities are
violated.

(vii) We consider the GUE as obtained by constraining the GSE. The
number of constraints is $N_Q = N(N-1)$, there are no systematic
degeneracies, and the Dyson index is $\beta = 4$. The
inequality~(\ref{mainbeta}) is violated, as it must be.

(viii) We consider an ensemble of real symmetric block--diagonal
matrices consisting of two blocks with equal dimensions.  Clearly,
eigenvalues belonging to different blocks are uncorrelated.  The
${\cal Q}$--space is given by the chiral GOE with quadratic blocks.
The number of constraints imposed on the GOE is $N_Q = N^2/4$, there
are no systematic degeneracies, and $\beta = 1$. The
inequality~(\ref{mainbeta}) is violated, as must be the case.

(ix) We consider the chiral GOE as obtained by constraining
the GOE. Matrices of the chiral GOE consist of two rectangular
off--diagonal blocks of size $n \times m$. For $m \ge n$, there are
$m-n$ zero eigenvalues, and the remaining $2n$ eigenvalues come in
$n$ pairs with opposite signs. The correlation of the positive
eigenvalues is determined by the expression~\cite{Ver94}
\be
\label{chgoe}
\prod_{1\le j<k\le n} |x_k^2-x_j^2| \prod_{l=1}^n x_l^{|m-n|} \ .
\ee
We consider two cases. First, we set $m = n$.  This case is converse
to example (viii), as the roles of ${\cal P}$-space and ${\cal
Q}$-space are interchanged. According to expression~(\ref{chgoe}),
eigenvalues with equal signs repel each other linearly, but there is
no repulsion between the smallest positive eigenvalue and the largest
negative eigenvalue. The number of constraints is $N_Q = n(n+1) = (N/2)
(N/2 +1)$, there are no systematic degeneracies, and $\beta = 1$. The
inequality~(\ref{mainbeta}) is violated, as it must be. For the second
case of the chiral GOE, we set $m = n+1$. In this case, the eigenvalue
spectrum has one zero eigenvalue, and $n$ pairs of eigenvalues with
opposite signs. We have $N_Q = (n+1)^2 = (N+1)^2/4$, there are no
systematic degeneracies, and $\beta = 1$. The inequality~(\ref{mainbeta})
is violated. However, inspection of expression~(\ref{chgoe}) shows
that there is linear level repulsion between any pair of neighboring
levels. In particular, the zero eigenvalue repels the smallest positive
eigenvalue. Again, we cannot draw any conclusion about level repulsion
from the inequality~(\ref{mainbeta}) since it is violated.

\section{Further properties of the CGUE}
\label{newsec}

Constraints affect not only the spectral fluctuations of the CGUE but
also the spectral width, and the distribution of matrix elements of
the CGUE. To show this for the spectral width, we recall the usual
normalization condition $\overline{H^2_{\mu \nu}} = \lambda^2/N$ of the
matrix elements $H_{\mu \nu}$ of the GUE. Here $N$ is the matrix
dimension, the overbar denotes the ensemble average, and $2 \lambda$ is
the radius of Wigner's semicircle. With this normalzation, the spectral
width of the GUE has the value $(1/N) {\rm Trace} \overline{H^2} =
\lambda^2$. To work out $(1/N) {\rm Trace} \overline{H^2}$ for the CGUE,
we start from the normalized distribution function of the eigenvalues,
\be
P_{\rm CGUE}(x) = {\cal N} \Delta^2(x) F(x) \exp( - \frac{N}{2 \lambda^2}
\sum_j x^2_j ) \ .
\label{width1}
\ee
Here ${\cal N}$ is a normalization factor. For $F(x) = 1$ (no constraints)
this function yields the GUE with the above--mentioned normalization. To
work out the normalization factor ${\cal N}$, we relate $P_{\rm CGUE}(x)$
to the distribution function $P_{\cal P}(x)$ defined in Eq.~(\ref{A8a})
with normalization factor ${\cal N}_0 = (2 \pi)^{(N_Q - N)/2} / \prod k!$.
We substitute $x_j = y_j \lambda / \sqrt{N}$ and use the fact that $F(x)
= (\sqrt{N}/\lambda)^{N_Q}$ to obtain
\be
{\cal N} = \big( \frac{N}{\lambda^2} \big)^{(N^2 - N_Q)/2} \ {\cal N}_0 \ .
\label{width2}
\ee
A straightforward calculation then yields
\be
\big( \frac{1}{N} {\rm Trace} \ \overline{H^2} \big)_{\rm CGUE} =
\lambda^2 [ 1 - \frac{N_Q}{N^2} ] \ .
\label{width3}
\ee
This shows that in comparison with the GUE, the spectral width of the
CGUE is reduced by the factor $\sqrt{1 - N_Q / N^2}$. Every constraint --
irrespective of whether it is located on the main diagonal or not, or
whether it affects the the real or the imaginary elements of $H$ --
yields the same contribution to the argument of the square root.

For the GUE, the matrix elements are Gaussian--distributed random
variables. Because of the constraints, this is not so for the CGUE. To
show this, we calculate the variance
\be
V_{\rm CGUE} = \overline{\bigg[ \frac{1}{N} \ {\rm Trace} \ H^2 \bigg]^2 }
- \bigg[ \frac{1}{N} \ {\rm Trace} \ \overline{H^2} \bigg]^2
\label{width4}
\ee
of $(1/N) {\rm Trace} \ H^2$ in two ways. Using $P_{\rm CGUE}(x)$ as given
by Eq.~(\ref{width1}) we obtain
\be
V_{\rm CGUE} = \frac{2 \lambda^4}{N^2} \big(1 - \frac{N_Q}{N^2}\big) \ .
\label{width5}
\ee
On the other hand, Eq.~(\ref{width3}) implies for the CGUE
\begin{equation} 
\overline{H_{\mu \nu} H_{\nu \mu}} = \frac{\lambda^2}{N} \big( 1 -
\frac{N_Q}{N^2} \big) \ .
\label{width6}
\end{equation}
Assuming now that the matrix elements do have a Gaussian distribution,
we find for the variance of $(1/N){\rm Trace} \ H^2$ the value
\begin{equation}
\frac{2 \lambda^4}{N^2} \big( 1 - \frac{N_Q}{N^2} \big)^2 \ .
\label{width7}
\end{equation}
This disagrees with Eq.~(\ref{width5}) and shows that the deviations
from the Gaussian distribution increase monotonically with increasing
number of constraints. While the matrix elements of the constrained
ensembles defined by Eq.~(\ref{A7a}) do have a Gaussian distribution,
this is not the case for their unitarily invariant counterparts defined
in Eqs.~(\ref{A8a}) and (\ref{A9a}).

\section{Complementary distribution functions}
\label{FT}

Equation~(\ref{A10}) expresses the joint probability function $P_{\cal
P}(x)$ of the CGUE as an integral over the constraints $\{B_q\}$. We
can interchange the roles of the ${\cal P}$--space and the ${\cal
Q}$--space and study the distribution function $P_{\cal Q}$ using the
set $\{B_p\}$ in ${\cal P}$--space as constraints. The probability
distribution $P_{\cal Q}(x)$ has the form of Eq.~(\ref{A8a}) with
\be
F_{\cal Q}(x) = \int {\rm d} [U] \bigg( \prod_p \delta( \langle
B_p | U x U^{\dagger} \rangle ) \bigg)
\label{C1a}
\ee
playing the role of $F_{\cal P}(x)$. The integrands of either function
depend only on the eigenvalues $b$ of the matrix $B$ of the
constraining ensemble. This suggests that it should be possible to
express $P_{\cal P}(x)$ as an integral over the complementary
distribution function $P_{\cal Q}(b)$, and vice versa. In this
Section we show that this is indeed the case and can be done via a
generalization of Fourier's theorem.

We start from the identity
\be
\fl
\qquad
\prod_q \delta( \langle B_q | H \rangle ) = {1\over (2 \pi)^{N_Q}}
\int \prod_q  {\rm d} t_q \int \prod_p  {\rm d} t_p 
\exp{\left(i \langle \sum_\sigma t_\sigma B_\sigma | H \rangle\right)} 
\prod_p \delta(t_p) \ .
\label{C1}
\ee
The matrices $B_\sigma$ form a basis of ${\cal V}$. Therefore, the
matrix $B = \sum_\sigma t_\sigma B_\sigma$ is the general Hermitean
matrix in $N$ dimensions. By a suitable orthogonal transformation, we
can replace the (general) basis matrices $B_\sigma$ by the special
basis set defined below Eq.~(\ref{compl}). For the transformed
integration variables $\tilde{t}_\sigma$, we have $\prod_\sigma {\rm d}
t_\sigma = \prod_\sigma {\rm d} \tilde{t}_\sigma$. The product
$\prod_\sigma {\rm d} \tilde{t}_\sigma$ is the product of the
differentials of all matrix elements and, hence, equal to ${\rm d} [B]$.
The delta functions $\delta(t_p)$ can be written identically as
$\delta(\langle B_p | B \rangle )$. Thus,
\be
\prod_q \delta( \langle B_q | H \rangle ) = {1\over (2 \pi)^{N_Q}}
\int {\rm d} [B] \ \exp{\left( i \langle B | H \rangle \right)} 
\prod_p \delta( \langle B_p |B \rangle) \ .
\label{C2}
\ee
The identity~(\ref{C2}) relates the constraints in ${\cal P}$--space
and in ${\cal Q}$--space via Fourier transformation. The Fourier
integral is taken over Hermitean matrices in $N$ dimensions.

We recall $H = U x U^{\dagger}$, write $B$ as $B = V b V^{\dagger}$,
and integrate the identity~(\ref{C2}) over the unitary group $U$, using
the definition~(\ref{A9a}) of $F_{\cal P}(x)$ and the HCIZ integral.
This yields
\ba
F_{\cal P}(x) &=& {1\over (2 \pi)^{N_Q}} \int {\rm d} [U] \int {\rm d}
[B] \ \exp{\left(i \langle B | U x U^\dagger \rangle\right)} 
\prod_p \delta( \langle B_p | B \rangle ) \nonumber \\
&=& {(2 \pi)^{N(N-1)/2-{N_Q}}\over \prod_{k=1}^N k!} 
\int {\rm d} [U] \int {\rm d} [V] \int {\rm d}
[b] \ \Delta^2(b) \ \exp{\left(i \langle V b V^{\dagger} | U x U^{\dagger}
\rangle \right)}\nonumber \\
&& \qquad \times \prod_p \delta( \langle B_p | VbV^\dagger \rangle )
\nonumber \\
&=& {(2 \pi)^{N(N-1)/2-{N_Q}}\over N! \ i^{N(N-1)/2}} 
 \int {\rm d} [b] \ \Delta(b) \ \frac{ \det [ \exp{( i b_j
x_k ) }]} {\Delta(x)} \nonumber \\
&& \qquad \times \int {\rm d} [V] \ \prod_p \delta( \langle B_p
| VbV^\dagger \rangle ) \ .
\label{C4}
\ea
Combining this result with the definition~(\ref{C1a}) of $F_{\cal
Q}(x)$, we obtain the identity
\be
\Delta(x) F_{\cal P}(x) =
{(2 \pi)^{N(N-1)/2-{N_Q}}\over N! \ i^{N(N-1)/2}}
\int {\rm d} [b] \ \det [ \exp{( i b_j x_l )}] \ \Delta(b) F_{\cal Q}(b) \ .
\label{C5}
\ee
This identity shows that $\Delta(x) F_{\cal P}(x)$ is the Fourier
transform of $\Delta(b)F_{\cal Q}(b)$. It is a generalization of
Fourier's theorem to a class of symmetric functions of the
variables $x_1, x_2, \ldots, x_N$. It is the main result of this
Section. We recall that $F_{\cal P}(x)$ and $F_{\cal Q}(b)$ are directly
related to the joint probability functions $P_{\cal P}(x)$ and
$P_{\cal Q}(b)$, respectively, via Eq.~(\ref{A8a}). Thus, the
complementary probability functions themselves are also obtained from
each other by a generalized Fourier transform.

Writing the identity~(\ref{C5}) for both $F_{\cal P}(X)$ and $F_{\cal
Q}(b)$, using the reality of $F_{\cal P}$ and $F_{\cal Q}$ in the
latter relation, and inserting the result into the former, we find
\ba
\fl
\qquad
\Delta(x) F_{\cal P}(x) = {(2\pi)^{-N}\over (N!)^2}
\int {\rm d}[b] \int {\rm d} [y] \det [ \exp{(i x_j b_l )}]
\det [ \exp{( - i b_j y_l )} ] \Delta(y) F_{\cal P}(y) \ . \nonumber \\
\label{C6}
\ea
Eq.~(\ref{C6}) implies that
\be
{(2\pi)^{-N}\over (N!)^2} \int {\rm d} [b] \det [ \exp{( i x_j b_l )} ] 
\det [ \exp{( - i b_j y_l )} ] = \prod_{l = 1}^N \delta(x_l - y_l)
\label{C7}
\ee
holds for a class of symmetric functions. 

To illuminate the result~(\ref{C5}), we discuss two examples. In the
first example, the ${\cal P}$--space consists of all real symmetric
matrices, and the ${\cal Q}$--space of all antisymmetric matrices with
purely imaginary matrix elements. The joint distribution function of
the GOE is known, and so is that of the (purely imaginary)
antisymmetric Gaussian ensemble~\cite{MehRos68}. Both these functions
are related via the generalized Fourier transform~(\ref{C5}), and the 
explicit calculation is presented in the Appendix. In the second
example, the ${\cal P}$--space consists of two block--diagonal matrices
with equal dimensions. Thus, the eigenvalue distribution is that of a
superposition of two GUEs.  The complementary ${\cal Q}$--space is
given by the chiral GUE used to describe universal phenomena of the QCD
Dirac operator. The eigenvalue distribution for this ensemble is also
known~\cite{Ver94}. Both eigenvalue distributions are related to each
other by the generalized Fourier transform. 

\section{Deformed Gaussian ensembles}
\label{deform}

So far, we have formulated the constraints in terms of delta functions 
(see Eq.~(\ref{A7a})). It is tempting to consider a more general case
where the matrices of the ${\cal P}$-- and the ${\cal Q}$--space both
have non-vanishing probability but differ in the widths of their
Gaussian distribution functions. By changing the width of the
distribution in ${\cal Q}$--space, we are able, for instance, to
study the transition from the GUE to the CGUE. We refer to such
ensembles as {\it deformed} Gaussian ensembles.

With $\{ B_\alpha \}$ an orthonormal basis in ${\cal V}$, we consider
an ensemble of Hermitean matrices
\be
H = \sum_{p} s_p B_p  + \sum_{q} s_q B_q \ ,
\label{gauss}
\ee
where $s_p$ and $s_q$ are real independent Gaussian variables with zero
mean and variances
\be
\overline{s^2_p} = \frac{1}{ \lambda_{\cal P}} \ , \
\overline{s^2_q} = \frac{1}{ \lambda_{\cal Q}} \ .
\ee
Here and in what follows, we indicate ensemble averages by a bar over
the quantity of interest. We note that for $\lambda_{\cal P} = 1$ and
$\lambda_{\cal Q} = \infty$, we recover the constrained
ensemble~(\ref{A7a}), while for $\lambda_{\cal P} = 1 = \lambda_{\cal
Q}$, we recover the GUE. We note that the deformed Gaussian
ensemble~(\ref{gauss}) is not unitarily invariant unless $\lambda_{\cal
P} = \lambda_{\cal Q}$. In calculating the spectral distribution
function, we do not proceed as in the main part of Section~\ref{CGUE},
where we constructed first the unitarily invariant ensemble. Rather, we
use the shortcut described below Eq.~(\ref{A9a}) and integrate directly
over the unitary group. This procedure automatically generates the
deformed Gaussian unitary ensemble (DGUE).

We make use of $s_\sigma = \langle H|B_\sigma\rangle$, 
${\rm d}[H]=\prod_\sigma {\rm d}s_\sigma$, and start from the probability 
density formula
\ba
\fl
\qquad
W_{\cal PQ}(H,\lambda_{\cal P},\lambda_{\cal Q}) {\rm d}[H]=\\ 
\left({\lambda_{\cal P}\over 2\pi}\right)^{N_P\over 2} 
\left({\lambda_{\cal Q}\over 2\pi}\right)^{N_Q\over 2} 
 \exp{\left(-{\lambda_{\cal P}\over 2} \sum_p \langle H|B_p\rangle^2 
-{\lambda_{\cal Q}\over 2}\sum_q \langle H|B_q\rangle^2\right)} {\rm d}[H] \ .\nonumber
\ea
We employ the completeness and orthonormality of $\{B_\sigma\}$, substitute
\be
\fl
\qquad
\lambda_{\cal P} \sum_p \langle H|B_p\rangle^2 
+\lambda_{\cal Q}\sum_q \langle H|B_q\rangle^2 = 
\lambda_{\cal P} \langle H|H\rangle 
+(\lambda_{\cal Q}-\lambda_{\cal P})\sum_q \langle H|B_q\rangle^2 \ ,
\ee
express $H$ in the diagonalized form, $H=UxU^\dagger$, and write ${\rm
d}[H]$ as in Eq.~(\ref{dh}). Inspecting the resulting formula, we find
that the distribution function $P_{\cal PQ}(x,\lambda_{\cal
P},\lambda_{\cal Q})$ of the eigenvalues of the DGUE is given by the
integral over the unitary group in $N$ dimensions
\ba
\fl
\qquad
P_{\cal PQ}(x,\lambda_{\cal P},\lambda_{\cal Q})
 = \left( {\lambda_{\cal P} \over 2\pi} \right)^{N_P\over 2} 
     \left( {\lambda_{\cal Q} \over 2\pi} \right)^{N_Q\over 2}
{(2\pi)^{N(N-1)/2} \over \prod_{k=1}^N k!}
\exp{\left( -{\lambda_{\cal P}\over 2} \langle x|x \rangle\right)}  
\Delta^2(x)\nonumber\\
 \times \int {\rm d} [U] 
\exp{\left(-{\lambda_{\cal Q}-\lambda_{\cal P}\over 2} \sum_q 
\langle UxU^\dagger |B_q \rangle^2 \right)} \ . 
\ea
We use the Hubbard-Stratonovich transformation
\ba
\exp{\left( - {\lambda_{\cal Q}-\lambda_{\cal P}\over 2} \langle UxU^\dagger
| B_q \rangle^2 \right)} = \nonumber\\ 
\qquad \sqrt{\lambda \over 2\pi} \int {\rm d}t_q \exp{ \left(i t_q \langle
UxU^\dagger |B_q \rangle - {\lambda\over 2} t^2_q \right)}
\label{hstrans}
\ea
where
\be
\label{tilde}
\lambda = { 1 \over \lambda_{\cal Q} -\lambda_{\cal P} }
\ee
and integrate over the unitary group with the help of the 
HCIZ formula~(\ref{hciz}). We obtain
\ba
\fl 
\qquad
P_{\cal PQ}(x,\lambda_{\cal P},\lambda_{\cal Q})
 = \left( {\lambda_{\cal P} \over 2\pi} \right)^{N_P\over 2} 
     \left( {\lambda_{\cal Q} \over 2\pi} \right)^{N_Q\over 2}
{(2\pi)^{N(N-1)/2} \over N! \ i^{N(N-1)/2}}
\exp{\left( -{\lambda_{\cal P}\over 2} \langle x|x \rangle\right)}  
\Delta^2(x)\nonumber\\
\times \left({\lambda \over 2\pi} \right)^{N_Q\over 2}
\int \prod_q {\rm d} t_q \ \exp \left( - {\lambda\over 2} t^2_q \right) 
{\det\left[ \exp{( i x_k b_l )} \right] \over \Delta(x) \Delta(b) } \ .
\label{probdis2}
\ea
The $\{ b_l \}$ are the eigenvalues of the matrix $B = \sum_q t_q
B_q$.  We note that for $\lambda_{\cal Q} \to \lambda_{\cal P} = 1$,
we have $\lambda \to \infty$, the Gaussian factors turn into delta
functions, the integral over the $\{t_q\}$ becomes a constant, and
$P_{\cal PQ}(x,\lambda_{\cal P},\lambda_{\cal Q})$ turns into the GUE
distribution. On the other hand, for $\lambda_{\cal P} = 1$ and
$\lambda_{\cal Q} \to \infty$, we have $\lambda \to 0$, the Gaussian
cutoff becomes irrelevant, and the right--hand side of
Eq.~(\ref{probdis2}) approaches that of Eq.~(\ref{A10}). Thus, the
DGUE correctly interpolates between the GUE and the CGUE.

We recall the discussion in Section~\ref{suff} and observe that the
integrals on the right--hand side of Eq.~(\ref{probdis2}) converge for
all positive values of $\lambda$. This shows that in the DGUE,
$P_{\cal PQ}(x,\lambda_{\cal P},\lambda_{\cal Q})$ vanishes whenever
two eigenvalues $x_k, x_l$ coincide no matter how large $\lambda_{\cal
Q}$, and level repulsion prevails. If level repulsion disappears, it
does so abruptly at $\lambda_{\cal Q} = \infty$. On physical grounds
we expect, of course, that as $\lambda_{\cal Q}$ increases, level
repulsion affects ever smaller distances between levels.

The last term on the right--hand side of Eq.~(\ref{probdis2}) can be
reinterpreted as the ensemble average of
$\det[\exp{(ix_jb_k)}]/(\Delta(x)\Delta(b))$ over the ${\cal
Q}$--space ensemble of matrices $B=\sum_qt_qB_q$ with
Gaussian-distributed $t_q$ with the variances $\overline{t_q^2} =
1/\lambda$. We have
\ba
\fl
\left({\lambda \over 2\pi} \right)^{N_Q\over2}
\int \prod_q {\rm d} t_q \ \exp \left( - {\lambda\over 2} t^2_q \right) 
{\det\left[ \exp{( i x_k b_l )} \right] \over \Delta(x) \Delta(b) } 
=
\int {\rm d}[b] P_{\cal Q}(b,\lambda)
{\det\left[ \exp{( i x_k b_l )} \right] \over \Delta(x) \Delta(b) }
\label{Z4}
\ea
where (cf. Eq.~(\ref{A8a}))
\ba
\fl
P_{\cal Q}(b,\lambda) = {(2\pi)^{N(N - 1)/2}\over \prod_{k = 1}^N k!}
\left({\lambda \over 2\pi} \right)^{N_Q\over 2}
\exp{\left( - {\lambda\over 2}\langle b | b \rangle \right)} \Delta^2(b)
\int {\rm d}[U] \prod_p\delta(\langle B_p|UbU^\dagger\rangle)
\ea
denotes the disribution function of eigenvalues $b_k$ of this ensemble. 
This shows that $P_{\cal PQ}(x,\lambda_{\cal P},\lambda_{\cal Q})$
can be expressed entirely through the rescaled eigenvalue distribution 
$P_{\cal Q}(b,\lambda)$:
\ba
\fl
\qquad
P_{\cal PQ}(x,\lambda_{\cal P},\lambda_{\cal Q})
 = \left( {\lambda_{\cal P} \over 2\pi} \right)^{N_P\over 2} 
     \left( {\lambda_{\cal Q} \over 2\pi} \right)^{N_Q\over 2}
{(2\pi)^{N(N-1)/2} \over N! \ i^{N(N-1)/2}}
\exp{\left( -{\lambda_{\cal P}\over 2} \langle x|x \rangle\right)}  
\Delta^2(x)\nonumber\\
\times\int {\rm d} [b] P_{\cal Q}(b,\lambda)  
{\det\left[ \exp{( i x_k b_l )} \right] \over \Delta(x) \Delta(b) } \ .
\ea

Throughout the derivations in this Section we may, of course, exchange
the roles of the ${\cal P}$--space and of the ${\cal Q}$--space. The
resulting equations are obtained by substituting $P \leftrightarrow Q$
except that the term in the exponent on the left--hand side of
Eq.~(\ref{hstrans}) is now positive. This causes the imaginary unit
$i$ to disappear in that and all following equations. We note that the
generalized Fourier transformation of Section~\ref{FT} can be extended
to the DGUE. We also note that the transition from GUE to GOE, or from
GUE to the antisymmetric Gaussian ensemble, can be described in the
framework of the DGUE. In this case, one recovers the result given in
Ref.~\cite{MehPan83}.

\section{Summary}

Introducing sets of constraints, we have defined new classes of
random--matrix ensembles, the constrained Gaussian unitary (CGUE), and
the deformed Gaussian unitary (DGUE) ensembles. The CGUEs consist of
Hermitean random matrices where the constraints set certain linear
combinations of matrix elements to zero. The DGUEs interpolate between
the GUE and the corresponding CGUE. Using the Harish--Chandra Itzykson
Zuber integral for the CGUE, we have found a sufficient condition for
GUE--type level repulsion to persist in the presence of
constraints. This condition depends only on the number of constraints
and on the internal symmetries of the constraining matrices and can be
formulated as a simple inequality. We have derived a generalized
Fourier theorem which relates the spectral properties of the
constraining ensembles with those of the constrained ones. We have
shown that in the DGUEs, level repulsion always prevails at
sufficiently short distances.  It is only in the limit of strictly
enforced constraints that level repulsion may be lifted. As a result,
we find that GUE--type level repulsion is remarkably robust. The
extension of this approach to the orthogonal and the symplectic cases
is hampered by the fact that the analogues of the Harish--Chandra
Itzykson Zuber integral formula are not known in closed form. In these
cases, we found a sufficient condition for the canonical level
repulsion only for certain classes of constraints.

\ack
This work was supported in part by the U.S. Department of Energy under
Contract Nos. DE-FG02-96ER40963 (University of Tennessee) and
DE-AC05-00OR22725 with UT-Battelle, LLC (Oak Ridge National
Laboratory).
Z.P. thanks the members of the Max-Planck-Institut f\"ur Kernphysik
in Heidelberg for their hospitality and support, and acknowledges support
by the Grant Agency of the Czech Republic (contract
No. 202/06/0363) in Prague.

\appendix
\setcounter{section}{1}

\section*{Appendix: Relation between the GOE and the antisymmetric 
Gaussian ensemble}
\label{AppA}

The Gaussian antisymmetric ensemble and the GOE are complementary
ensembles. In this appendix, we show that the eigenvalue distributions
of GOE and the antisymmetric Gaussian ensemble are related to each
other by the generalized Fourier transform~(\ref{C5}). For simplicity,
we consider matrix ensembles with matrices of even dimension $N = 2m$.
We start by considering the GOE basis matrices as the constraints and
use the probability distribution of the antisymmetric Gaussian ensemble
as given in Ref.~\cite{MehRos68}
\be
\label{anti}
P_{\cal P}(b) \propto {\cal S} \exp{\left(-{1\over 2} 
\langle b|b\rangle\right)} \prod_{1\le j<k\le m} (b_j^2-b_k^2)^2 
\prod_{l=1}^m \delta(b_l + b_{m+l}) \ .
\ee
We have omitted an overall normalization constant. The symbol ${\cal
S}$ denotes symmetrization with respect to all eigenvalues. According
to Eq.~(\ref{A8a}), one thus finds
\ba
\label{fp}
\Delta(b) F_{\cal P}(b) &\propto& {\cal A} 
\prod_{1\le j<k\le m} (b_j^2-b_k^2)^2 
\prod_{l=1}^m \delta(b_l + b_{m+l}) / \Delta(b) \\\nonumber
&\propto& {\cal A} \prod_{k=1}^m {\delta(b_k + b_{m+k})\over b_k} \ .
\ea
The symbol ${\cal A}$ denotes antisymmetrization with respect to all
eigenvalues. We have used the identity
\be
\label{appid}
\prod_{1\le k<l\le m} (x_k^2-x_l^2)^2 \prod_{j=1}^m \delta(x_j+x_{j+m})
 = \Delta(x) \prod_{j=1}^m {\delta(x_j+x_{j+m}) \over x_j - x_{j+m}} \ ,
\ee
which can proved by expanding the products and exploiting the 
$\delta$-functions. We insert the last relation~(\ref{fp}) into the
Fourier relation~(\ref{C5}) and find
\ba
\Delta(x) F_{\cal Q}(x) \propto \int \prod_{l=1}^m {{\rm d} b_l\over b_l} 
{\rm det}\left[\exp{(i b_j x_k)}\right] \ . 
\ea
We have performed one-half of the integrations by means of the
$\delta$--functions, and $b_{m+j} = -b_j$ thus holds in the exponential.
We have dropped the antisymmetrizer ${\cal A}$ since the determinant is
antisymmetric anyway. We use the Laplace expansion of the determinant and
obtain terms of the form
\be
\prod_{k=1}^m \int {{\rm d} b_k\over b_k}  
\exp{\left(i b_k(x_{j_k}-x_{j_{m+k}})\right)} = (i\pi)^m
\prod_{k=1}^m {\rm sign} (x_{j_k}-x_{j_{m+k}}) \ .
\ee
Summing up all terms, one finds
\be
\Delta(x) F_{\cal Q}(x) \propto {\cal A} \prod_{k=1}^m 
{\rm sign} (x_{j_k}-x_{j_{m+k}}) \ .
\ee
This expression is totally antisymmetric under permutations of any two
eigenvalues and is a homogeneous function of degree zero. Thus,
\be
\label{Fgoe}
\Delta(x) F_{\cal Q}(x) \propto {\rm sign} (\Delta(x)) \ ,
\ee
and this yields the GOE distribution once Eq.~(\ref{A8a}) is employed.

We can also employ the Fourier transform to go from the GOE to the
antisymmetric Gaussian ensemble. For this purpose, we insert the GOE
expression~(\ref{Fgoe}) into the Fourier relation~(\ref{C5}) and find
\be
\Delta(x) F_{\cal P}(x) \propto \int {\rm d} [b] \, {\rm sign} (\Delta(x))
{\rm det}\left[\exp{(i b_j x_k)}\right] \ .
\ee
This integral can be performed by using, for instance, the results
presented in Sect. 3 of Ref.~\cite{MehPan83}. We find
\be
\label{app1a}
\Delta(x) F_{\cal P}(x) \propto {\rm Pf}[a_{jk}]_{j,k=1\ldots2m} \ . 
\ee
Here, 
\ba
a_{jk} &\equiv& \int\limits_{u<v} {\rm d} u {\rm d} v 
\left( e^{ix_j u} e^{i x_k v} -  
e^{ix_k u} e^{i x_j v}\right) \\\nonumber
&=& 8\pi i (x_k-x_j)^{-1} \delta(x_j+x_k) \ , 
\ea
and ${\rm Pf}$ denotes the Pfaffian 
\be
{\rm Pf}[a_{jk}] = \sum_P (-1)^P a_{j_1 j_2} a_{j_3 j_4}\ldots
a_{j_{2m_1} j_{2m}} \ ,
\ee
with $P$ running over all $(2m)! /(2^m m!)$ permutations $(j_1,\ldots,
j_{2m})$ of $(1,2,\ldots, 2m)$ with restrictions $j_1 < j_2, j_3 < j_4,
\ldots j_{2m_1} < j_{2m}$ and $i_1 < i_3 < \ldots i_{2m-1}$. 

The expression~(\ref{app1a}) is thus completely antisymmetric under
permutations of any two of its arguments. The probability density we seek
is essentially this expression, multiplied by the Vandermonde
determinant $\Delta(x)$, and is thus completely symmetric under
permutations of its arguments. We may therefore focus on just one term
of the Pfaffian and write
\be
P_{\cal P}(x) \propto {\cal S} 
\exp{\left(-{1\over 2}\langle x|x\rangle\right)} \Delta(x) 
\prod_{k=1}^m (x_{k} - x_{k+m})^{-1} \delta(x_{k} + x_{k+m}) \ . 
\ee
Employing the $\delta$-functions and the identity~(\ref{appid}), we recover
Eq.~(\ref{anti}).

\end{document}